\documentclass[aps,pra,twocolumn,epsfig,superscriptaddress,showpacs,amsmath,amsfonts,amssymb,floatfix,longbibliography]{revtex4-1}
\usepackage{amsmath,amsfonts,amssymb,color}
\usepackage[hidelinks]{hyperref}
\usepackage{graphicx}
\usepackage{dcolumn}
\usepackage{soul}
\usepackage{bm}

\begin{document}
\title{Non-Hermitian dynamics of slowly-varying Hamiltonians }

\author{Hailong Wang}
\email{hailong.china@gmail.com}
\affiliation{Division of Physics and Applied Physics, School of Physical and Mathematical Sciences, Nanyang Technological University, Singapore 637371, Singapore}

\author{Li-Jun Lang}
\affiliation{Division of Physics and Applied Physics, School of Physical and Mathematical Sciences, Nanyang Technological University, Singapore 637371, Singapore}

\author{Y.~D.~Chong}
\email{yidong@ntu.edu.sg}
\affiliation{Division of Physics and Applied Physics, School of Physical and Mathematical Sciences, Nanyang Technological University, Singapore 637371, Singapore}
\affiliation{Centre for Disruptive Photonic Technologies, Nanyang Technological University, Singapore 637371, Singapore}

\date{ \today}

%%%%%%%%%%%%%%%%%%%% ABSTRACT %%%%%%%%%%%%%%%%%%%%
\begin{abstract}

We develop a theoretical description of non-Hermitian time evolution that accounts for the breakdown of the adiabatic theorem.  We obtain closed-form expressions for the time-dependent state amplitudes, involving the complex eigenenergies as well as inter-band Berry connections calculated using basis sets from appropriately-chosen Schur decompositions.  Using a two-level system as an example, we show that our theory accurately captures the phenomenon of ``sudden transitions'', where the system state abruptly jumps from one eigenstate to another.

\end{abstract}
%%%%%%%%%%%%%%%%%%%%%%%%%%%%%%%%%%%%%%%%%%%%%%%%%%

\maketitle

\section{Introduction}

The dynamical features of non-Hermitian systems have attracted a great deal of recent interest~\cite{de2006time, bender2007making, mostafazadeh2010pseudo, gong2013time, milburn2015general, gong2015stabilizing, gong2016aharonov}, driven in large part by the field of photonics~\cite{el2007theory, klaiman2008visualization, guo2009observation, ruter2010observation, longhi2015robust}, where non-Hermiticity (in the form of optical gain and/or loss) can be easily introduced and controlled.  Researchers have uncovered a variety of phenomena tied intrinsically to non-Hermiticity, including laser-absorbers~\cite{longhi2010pt, chong2011p, wong2016lasing}, unidirectional light transport~\cite{lin2011unidirectional, peng2014parity}, asymmetric mode conversion~\cite{ghosh2016exceptional, doppler2016dynamically, xu2016topological, feng2013experimental, hassan2017dynamically, hassan2017chiral}, and exceptional point-aided sensing~\cite{chen2017exceptional, hodaei2017enhanced}.  Alongside these efforts, considerable theoretical work has gone into understanding the distinctive features of non-Hermitian dynamics~\cite{bender1998real, bender2002complex, moiseyev2011non, khantoul2017invariant, maamache2017non, zhou2017dynamical, longhi2017oscillating, longhi2017floquet, yip2018quantum}.

Common analytic methods developed for Hermitian dynamical systems, including time-dependent perturbation theory (for weakly-perturbed Hamiltonians) and adiabatic theory (for slowly-changing Hamiltonians)~\cite{kato2013perturbation}, tend to be either fundamentally inapplicable or poorly performing for non-Hermitian systems~\cite{nenciu1992adiabatic, sun1993high, mostafazadeh2014adiabatic, longhi2017non, longhi2017oscillating}.  The usual time-dependent perturbation theory breaks down because the eigenstates of a non-Hermitian Hamiltonian are not orthogonal, which can make the expansion of a perturbed time evolution operator very sensitive to initial conditions~\cite{zyablovsky2016parametric}; another related problem is that perturbative corrections involve transitions between eigenstates, and in a non-Hermitian system the corresponding small state amplitudes can undergo exponential growth relative to the rest of the state vector.  For similar reasons, the adiabatic theorem does not apply to systems with slowly-varying non-Hermitian Hamiltonians~\cite{berry2011slow, ibanez2014adiabaticity, milburn2015general}.

In this paper, we develop an analytic method to describe non-Hermitian time evolution, applicable to slowly-varying time-dependent non-Hermitian Hamiltonians.  We introduce several sets of orthonormal basis states produced by different Schur decompositions (one for each eigenstate/eigenvector), and use them to derive closed-form expressions for the state amplitudes.  These expressions involve the complex eigenenergies as well as inter-band Berry connections calculated using the chosen basis vectors (rather than bi-orthogonal products), and they use the fastest-amplifying (or slowest-decaying) eigenenergy as a natural reference scale factor.  The results can be regarded as generalizing earlier descriptions of Hermitian dynamics in terms of adiabatic evolution and higher-order corrections to adiabaticity~\cite{berry1984, wang2015}.  We then show, using numerical examples, that our theory accurately describes the phenomenon of ``sudden transitions'', in which a non-Hermitian system's state abruptly jumps from following one eigenstate to following a different eigenstate (the resulting ``asymmetric mode conversion'' behavior has recently demonstrated in microwave~\cite{doppler2016dynamically}, optomechanical~\cite{xu2016topological}, and optical~\cite{feng2013experimental} experiments).  We find that the key role in the sudden transitions is played by a set of functions originating from effective inter-band hoppings.

The manuscript is organized as follows.  In Section~\ref{sec:perturbation}, we show how Hermitian evolution can be expressed in terms of systematic corrections to adiabaticity, by deriving integral expressions for the state amplitudes involving the eigenenergies and inter-band Berry connections.  In section~\ref{sec:Schur}, we discuss the Schur decompositions of non-Hermitian matrices, and introduce a set of Schur decompositions (and their associated orthonormal basis) suitable for keeping track of non-Hermitian evolution.  In Section~\ref{sec:nonhermitian}, we derive closed-form expressions for the state amplitudes of two-level as well as higher-dimensional non-Hermitian systems.  In Section~\ref{sec:Model}, we subject the theory to numerical tests.  We conclude in Section~\ref{sec:conclu}.

\section{Non-adiabatic corrections to Hermitian evolution}
\label{sec:perturbation}

We begin by revisiting the evolution of Hermitian systems, identifying how corrections to adiabaticity can be systematically accounted for, and observing how the features of Hermitian evolution break down for non-Hermitian Hamiltonians.

Let $H(t)$ be a time-dependent Hermitian Hamiltonian.  For convenience, we define a scaled time $s\equiv t/T$, where $T$ is a characteristic time duration along the ``trajectory'' of the Hamiltonian in some parameter space.  Taking $\hbar=1$, the time-dependent Schr\"odinger equation is 
\begin{equation}
  i\,T\,\frac{\partial}{\partial s}\,|\Psi(s)\rangle=H(s)\,|\Psi(s)\rangle\;.
  \label{eqn:schr}
\end{equation}
The state of the system can be expanded as a superposition of instantaneous eigenstates,
\begin{equation}
  |\Psi(s)\rangle = \sum_j c_j(s)\,e^{-i\Omega_j(s)}\,|\psi_j(s)\rangle\;,
  \label{eqn:state}
\end{equation} 
where $c_j(s)$ is the complex quantum amplitude for state $j$ at instant $s$, $\Omega_j(s)=T\int_0^s ds'\,E_j(s')$ is the accumulated dynamical phase, and $E_j(s)$ is the instantaneous eigenenergy.  We take the initial time to be $s = 0$.  Substituting Eq.~\eqref{eqn:state} into Eq.~\eqref{eqn:schr}, and projecting it onto $\langle\psi_k(s)|$, yields
\begin{equation}
  \dot{c}_k = i\,\sum_{j\ne k} c_j\, e^{i(\Omega_k-\Omega_j)}\, \mathcal{A}_{kj}.
  \label{eqn:diff1}
\end{equation} 
Here, $\dot{c}_k \equiv dc_k/ds$, and $\mathcal{A}_{kj} \equiv i\langle\psi_k| (d/ds) |\psi_j\rangle$ is the non-Abelian Berry connection~\cite{yu2011equivalent, gao2014field}, which describes the off-diagonal (``inter-band'') couplings between $|\psi_j\rangle$ and $|\psi_k\rangle$.  We adopt the parallel transport gauge in which the intra-band connection vanishes: $\langle\psi_j| (d/ds) |\psi_j\rangle = 0$.  (This means that if the trajectory forms a closed loop in parameter space, the basis functions for each point in parameter space may be different during different cycles~\cite{berry1984}, but that is not a problem for us.)

In standard time-dependent perturbation theory, the next step is to re-write Eq.~\eqref{eqn:diff1} in integral form and develop it into a Dyson series.  We instead follow a method, adopted from Ref.~\onlinecite{wang2015}, that allows us to distinguish between the ``degree of non-adiabaticity'' of various contributions to the state evolution; this will be helpful for making contact with the non-Hermitian theory later.  Let us define
\begin{equation}
  \mathcal{U}_{kj}(s) \equiv\exp\Big\{ i\big[ \Omega_k(s)-\Omega_j(s) \big]\Big\}.
\end{equation}
Note that $d\,\mathcal{U}_{kj}/ds = iT\left( E_k-E_j \right) \, \mathcal{U}_{kj}$.  We use this to integrate Eq.~\eqref{eqn:diff1}, obtaining
\begin{equation}
  \Delta c_k(s) = \sum_{j\ne k} \int_0^s ds' \;c_j \, \dot{\mathcal{U}}_{kj}\,
  \frac{\mathcal{A}_{kj}}{T\left(E_k-E_j\right)},
  \label{eqn:cks}
\end{equation}
where $\Delta c_k(s) \equiv c_k(s) - c_k(0)$.

Next, define
\begin{equation}
  \rho_{kj}^{(1)}(s) \equiv \frac{\mathcal{A}_{kj}(s)}{T\left[E_k(s) - E_j(s)\right]}.
  \label{eqn:rho1}
\end{equation}
This quantity, which involves both the non-Abelian Berry connection and the band energies, governs the non-adiabatic corrections to state $k$ induced by state $j$.  Using it, we integrate Eq.~\eqref{eqn:cks} by parts to obtain
\begin{multline}
  \Delta c_k(s) = \sum_{j\ne k} \Bigg\{ 
    \left[ c_j\, \rho^{(1)}_{kj} \, \mathcal{U}_{kj} \right]^s_{0} \\
    -i \int_0^s ds'\;c_k\,\rho^{(1)}_{kj}\, \mathcal{A}_{jk}+\Theta_{kj}\Bigg\} \;,
  \label{eqn:integ1}
\end{multline}
where
\begin{equation}
  \Theta_{kj} = - \int_0^s ds' \; c_j\, \mathcal{U}_{kj}
  \left( \dot\rho^{(1)}_{kj} +i\,\sum_{\ell\ne k} 
  \rho^{(1)}_{k\ell} \, \mathcal{A}_{\ell j} \right) \;.
  \label{eqn:thetakj}
\end{equation}
Note that no approximations have been made so far. The results up to Eq.~\eqref{eqn:thetakj} were previously derived in Ref.~\onlinecite{wang2015}, in order to study the corrections to adiabaticity in periodically driven Hermitian systems. There, the residual $\Theta_{kj}$ was dropped in order to calculate the lowest-order corrections (the resulting theory was successfully demonstrated in a recent experiment \cite{ma2017experimental}). 

Here, we show that the residual $\Theta_{kj}$ can be systematically accounted rather than being dropped, which will be useful for handling the non-Hermitian case. Let us define
\begin{align}
  \rho_{kj} & \equiv \rho_{kj}^{(1)} + \rho_{kj}^{(2)} + \rho_{kj}^{(3)} + \cdots \label{eqn:rhoseries} \\
  \rho_{kj}^{(n)} & \equiv \frac{i}{T\left(E_k-E_j\right)} \left[ \dot\rho_{kj}^{\,(n-1)} 
    + i\,\sum\limits_{\ell\ne k} \rho_{k\ell}^{(n-1)} \mathcal{A}_{\ell j} \right]. \label{taun}
\end{align}
Then the residual in Eq.~\eqref{eqn:integ1} can be absorbed into the other terms, producing the result
\begin{equation}
  \Delta c_k(s) =\sum_{j\ne k}
    \Big[ c_j \rho_{kj}\, \mathcal{U}_{kj} \Big]^s_{0} 
    - i\int_0^sds'\,c_k \sum_{j\ne k} \rho_{kj}\, \mathcal{A}_{jk}.
  \label{eqn:integ2}
\end{equation}
We call $\rho_{kj}$ the ``inter-band coherence factor'', and it describes the summed contributions of band $j$ to the non-adiabatic dynamics of band $k$.  In the adiabatic limit ($T\rightarrow \infty$), $\rho_{kj} \rightarrow 0$, and hence the entire right side of Eq.~\eqref{eqn:integ2} vanishes.

The first term on the right side of Eq.~\eqref{eqn:integ2} describes an effective \textit{inter-band} hopping from $j$ to $k$, evaluated solely at the initial and final times.  The second term can be regarded as an effective \textit{intra-band} contribution, since it involves the amplitudes of the same band $k$ at different points along the trajectory; however, this contribution is modulated by the non-Abelian Berry connections and inter-band coherence factors coming from all other bands.  Later, we will re-visit the significance of these features, in the non-Hermitian context.

Computationally speaking, it is \textit{not} necessarily advantageous to calculate $c_k(s)$ using Eq.~\eqref{eqn:integ2}.  Such a calculation requires diagonalizing the Hamiltonian and computing the inter-band Berry connections at each time step, which is not apparently any easier than directly integrating the Schr\"odinger equation.  The significance of Eq.~\eqref{eqn:integ2} is that it provides a description of how the state evolves, expressed in terms of a minimal set of quantities derived from the Hamiltonian.  This follows the spirit of Berry's theory of adiabatic quantum processes~\cite{berry1984}, in which state evolution was described in terms of the \textit{intra-band} Berry connection.  In the present case, the quantities of interest are the inter-band Berry connections and inter-band coherence factors.

Unlike the Dyson series formulation of time-dependent perturbation theory, which is based on the weakness of the time-dependent part of the Hamiltonian relative to its time-independent part, the present formulation relies on the time variation being slow relative to the characteristic energy level spacings.  For finite $T$, each consecutive term in Eq.~\eqref{eqn:rhoseries} represents a higher-order non-adiabatic correction to the inter-band coherence factor.  In order for the series to converge, so that the resummation leading to Eq.~\eqref{eqn:integ2} is valid, we need $|\dot\rho_{kj}^{(n)}/\rho_{kj}^{(n)}|\ll T|E_k-E_j|$, which is a weaker requirement than the usual adiabatic criterion $|\rho_{kj}^{(1)}|\ll 1$~\cite{tong2007sufficiency}.  Eq.~\eqref{eqn:integ2} can then be used to calculate $c_k(s)$ iteratively.  As we shall see, a variant of this derivation holds in the non-Hermitian case.

State evolution for non-Hermitian Hamiltonians differs from the Hermitian case in two important ways~\cite{Heiss1999, Dembowski2001}.  First, the eigenvalues of a non-Hermitian Hamiltonian are usually complex, with the imaginary part corresponding to an amplification rate (if positive) or decay rate (if negative).  The relative exponential growth of some states relative to others exacerbates the breakdown of adiabaticity; we can see in Eq.~\eqref{eqn:integ2} that if the $\mathcal{U}_{kj}$ factors grow exponentially, transitions between many different basis states become non-negligible.  Second, the eigenstates of a non-Hermitian Hamiltonian are usually not orthogonal to each other, which requires a modification in the derivation around Eq.~\eqref{eqn:diff1}.  Near an exceptional point (EP) of the Hamiltonian, several (usually two) eigenstates coalesce to become linearly dependent~\cite{Heiss1999, Dembowski2001}, and it is common for the state amplitudes to change drastically during time evolution~\cite{uzdin2011observability, berry2011slow, heiss2016mathematical, doppler2016dynamically, xu2016topological}.  One of the main objectives of this paper is to cast the equations for non-Hermitian evolution into a form where these problematic features can be kept under control.

\section{Schur decompositions}
\label{sec:Schur}

Our approach to non-Hermitian dynamics is based on Schur
decompositions.  A non-Hermitian Hamiltonian $H$ typically lacks an
orthogonal set of eigenvectors, but we can always find a unitary
matrix $U$ such that
\begin{equation}
  A=U^\dagger HU
\end{equation}
is upper triangular, with the eigenvalues of $H$ appearing along its diagonal.  The matrix $A$ is called a Schur form of $H$.  Unlike diagonalization, the Schur decomposition of a non-Hermitian matrix has a very useful feature: the columns of $U$ always form a complete orthonormal basis.  Moreover, it can be shown that for each eigenvector $x$ and eigenvalue $\lambda$ for $A$, there is a corresponding eigenvector $Ux$ and eigenvalue $\lambda$ for $H$.  This implies that the first column of $U$ is an eigenvector of $H$ (this is however not generally true for the other basis vectors).

The Schur decomposition is not unique.  In particular, we can apply a transformation to re-arrange the diagonal entries of the Schur form $A$ to any desired order.  Each choice of Schur decomposition produces a different orthonormal basis (column vectors of $U$).

Suppose the eigenvalues of $H$ are non-degenerate and denoted by $\lambda_j$ ($j=1,2,\dots,n$).  For our purposes, it is convenient to define a ``growth-ordered Schur decomposition'' that arranges the eigenvalues in \textit{descending} order of $\textrm{Im}[\lambda_j]$ along the diagonal of $A_1$.  In physical terms, this means listing the most amplifying (or least decaying) state first, followed by states of decreasing amplification rate.  Let $U_1$ be the transformation matrix for the growth-ordered Schur decomposition, and let its column vectors be $\{\,|\chi_1\rangle, |\chi_2\rangle, \cdots, |\chi_n\rangle\, \}$; the first one, $|\chi_1\rangle$, is also the most amplifying eigenvector of $H$.  We can represent $H$ as a sum of orthogonal projectors,
\begin{equation}
  H \,=\, \sum_i \lambda_i \, |\chi_i\rangle\langle\chi_i|
  \,+\, \sum_{i<j} {C_{ij} |\chi_i\rangle\langle\chi_j|}.
  \label{eqn:SchurH}
\end{equation}

Starting from the growth-ordered Schur decomposition $A_1$, we can generate another Schur form where the diagonal entries are $\lambda_2,\lambda_1,\lambda_{3},\dots,\lambda_n$ (i.e., with the second most amplifying state moved to the front).  The required transformation has the form
\begin{equation}
  A_2 = U_2^\dagger A_1 U_2, \quad U_2= 
  \begin{pmatrix}
  W^{(2)} & 0 \\ 0 & I_{n-2} \\
  \end{pmatrix},
\end{equation}
where $W^{(2)}$ is a $2\times2$ unitary matrix and $I_{n-2}$ is an identity matrix of size $(n-2)\times(n-2)$.  In the basis defined by this new Schur decomposition, the first basis vector (i.e., the first column of $U_1U_2$) is
\begin{equation}
  |\xi_2\rangle \equiv \Big[|\chi_1\rangle,|\chi_2\rangle\Big] \cdot W^{(2)}_{\bullet,1} \;,
\end{equation}
where $W^{(2)}_{\bullet,1}$ denotes the first column of $W^{(2)}$.  This basis vector is an eigenvector of $H$ with eigenvalue $\lambda_2$.  Moreover, we denote the \textit{second} basis vector in the basis by
\begin{equation}
  |\eta_2\rangle \equiv
  \Big[|\chi_1\rangle,|\chi_2\rangle\Big] \cdot\, W^{(j)}_{\bullet,2} \;.
\end{equation}
Note that this basis vector is ``associated'' with $\lambda_1$, but it is \textit{not} an eigenvector of $H$.

A similar swapping procedure can be performed to move the $j$-th eigenvalue ($j = 2, \dots, n$) to the front, so that the diagonal entries in the Schur form are $\lambda_j,\lambda_1,\dots,\lambda_{j-1},\lambda_{j+1},\dots$~(note that we keep the rest of the list in the same order; in particular, $\lambda_1$ is second).  Details of the procedure can be found in Appendix~\ref{Utransformation.apx}.  Each such Schur decomposition defines an orthonormal basis; we let $|\xi_j\rangle$ denote the first basis vector, and $|\eta_j\rangle$ denote the second basis vector.

In this way, we arrive at two sets of vectors, $\{|\xi_2\rangle, \dots, |\xi_n\rangle \}$ and $\{|\eta_2\rangle, \dots, |\eta_n\rangle \}$.  Within each set, the vectors  are not generally orthogonal.  Each $|\xi_j\rangle$ is an eigenvector of $H$, with eigenvalue $\lambda_j$.  Each $|\eta_j\rangle$ is associated with $\lambda_1$ in the particular Schur decomposition where $|\xi_j\rangle$ is the first column vector, and is orthogonal to $|\xi_j\rangle$.  As a generalization of Eq.~\eqref{eqn:SchurH}, for each $j$ we can decompose $H$ as
\begin{equation}
  H = \lambda_j \, |\xi_j\rangle\langle\xi_j|
  + \lambda_1 |\eta_j\rangle\langle\eta_j|
  + C_j |\xi_j\rangle \langle \eta_j|
  + \cdots \;,
  \label{eqn:SchurH2}
\end{equation}
where $C_j$ is some coupling coefficient, and the omitted terms involve kets orthogonal to both $|\xi_j\rangle$ and $|\eta_j\rangle$.

It can be shown that the set $\{|\chi_1\rangle,|\eta_2\rangle,\dots,|\eta_n\rangle\}$ is complete, provided that $|\chi_1\rangle$ is non-orthogonal to all the other eigenvectors $|\xi_{j}\rangle$ ($j = 2, \dots, n$); for details, see Appendix~\ref{completeness.apx}.  We shall argue that this basis set has the right properties for coping with the pathological features of non-Hermitian time evolution.  

\section{Non-Hermitian evolution}
\label{sec:nonhermitian}

We now adopt the following strategy for describing non-Hermitian evolution: First, we use a succession of Schur decompositions to construct a basis set $\{|\chi_1\rangle,|\eta_2\rangle,\dots,|\eta_n\rangle\}$ that is non-orthogonal but complete, as described in Section~\ref{sec:Schur}.  We use this to decompose the initial state vector, and follow the dynamics of each component separately.  The first component is handled by using the growth-ordered Schur decomposition to write the Hamiltonian in the form \eqref{eqn:SchurH}.  Since $|\chi_1\rangle$ is the most amplifying eigenstate, this component of the system state ``clings'' to it, in a manner similar to adiabatic following~\cite{berry2011slow, uzdin2011observability}.  However, the $C_{ij}$ coefficients in Eq.~\eqref{eqn:SchurH} are generally nonzero, and describe amplitude transfers from other bands to the most amplifying band.  This is accounted for by a formalism similar to the Hermitian case described in Section~\ref{sec:perturbation}, with a non-Hermitian variant of the inter-band coherence factor.

For each of the other components of the system state, the time evolution can be obtained from Eq.~\eqref{eqn:SchurH2}, which describes a coupling from $|\eta_j\rangle$ to either $|\eta_j\rangle$ or $|\xi_j\rangle$, \textit{and no other state}.  Since $|\eta_j\rangle$ and $|\xi_j\rangle$ are orthogonal, their dynamics can likewise be handled similarly to Section~\ref{sec:perturbation}.

\subsection{$2\times2$ non-Hermitian Hamiltonian}
\label{2by2}

We now work through the above process for the simple case of a $2\times2$ non-Hermitian Hamiltonian $H$.  Let the eigenvalues of $H$ be
$\lambda_1$ and $\lambda_2$, with $\mathrm{Im}(\lambda_1) >
\mathrm{Im}(\lambda_2)$.  We can perform two different Schur
decompositions.  In the notation of Section~\ref{sec:Schur},
\begin{align}
  \begin{aligned}
    H &= \lambda_1|\chi_1\rangle\langle\chi_1|
    + \lambda_2|\chi_2\rangle\langle\chi_2|
    + C_1 |\chi_1\rangle\langle\chi_2| \\
    &=\lambda_2|\,\xi_2\rangle\langle\,\xi_2|
    + \lambda_1|\eta_2\,\rangle\langle\,\eta_2|
    + C_2 |\xi_2\,\rangle \langle\,\eta_2|,
  \end{aligned}
\end{align}
where $\{|\chi_1\rangle, |\chi_2\rangle\}$ and $\{|\xi_2\rangle, |\eta_2\rangle\}$ are two distinct orthonormal bases.  When $H$ is time-dependent, these bases are likewise time-dependent.  Similar to Section~\ref{sec:perturbation}, we define a scaled time $s\equiv t/T$ and adopt the parallel-transport gauge for all bands (e.g., $\langle\chi_1|\dot\chi_1\rangle=0$).

We now use $\{|\chi_1(0)\rangle, |\eta_2(0)\rangle\}$ to decompose the initial system state:
\begin{equation}
  |\Psi(0)\rangle = a_1(0)|\chi_1(0)\rangle + b_2(0) |\eta_2(0)\rangle \;.
  \label{eqn:psi_0}
\end{equation}
Since the Schr\"odinger equation is linear, the dynamics of these two components can be handled separately.

Consider the first component.  Let $|\Psi_1(s)\rangle$ denote this part of the system state at each (rescaled) time $s$.  We project it onto the orthonormal basis $\{|\chi_1(s)\rangle, |\chi_2(s)\rangle\}$:
\begin{equation}
  |\Psi_1(s)\rangle = a_1(s) e^{-i\Omega_1} |\chi_1(s)\rangle
  + a_2(s) e^{-i\Omega_2}|\chi_2(s)\rangle,
  \label{eqn:psi_a}
\end{equation}
where $\Omega_j(s)\equiv T\int_0^s ds'\, \lambda_j(s')$.  Substituting this into the time-dependent Schr\"odinger equation, and left-multiplying by $\langle\chi_1(s)|$ and $\langle\chi_2(s)|$, yields
\begin{align}
  \dot{a}_1 & = i\, \mathcal{U}_{12}
  \Big(\mathcal{A}_{12}^a - T C_1\Big)\, a_2 \label{eqn:diff_a1} \\ 
  \dot{a}_2 &= i\, \mathcal{U}_{21} \, \mathcal{A}_{21}^a \,a_1,
  \label{eqn:diff_a2}
\end{align}
where
\begin{align}
  \mathcal{U}_{kj}(s) &\equiv \exp \{ i [\Omega_k(s)-\Omega_j(s) ]\} \\
  \mathcal{A}_{kj}^a(s) &\equiv i\langle \chi_k|\dot{\chi}_j \rangle. \label{berry}
\end{align}
In Eq.~\eqref{eqn:diff_a1}, the coupling from $|\chi_2\rangle$ to $|\chi_1\rangle$ involves the combination $\mathcal{A}_{12}^a - T C_1$.  Comparing this to Eq.~\eqref{eqn:diff1}, we see that the Berry connection $\mathcal{A}_{12}^a$ can be interpreted as a non-adiabatic but Hermitian-like contribution, while the $TC_1$ part is a purely non-Hermitian contribution involving a Schur coefficient (i.e., a non-diagonal component of the Schur form).

Note also that the non-Abelian connection \eqref{berry} is calculated with the orthonormal basis $\{|\chi_1\rangle, |\chi_2\rangle, \cdots\}$, using the usual inner product, similar to the non-Abelian Berry connection for Hermitian systems.  It is \textit{not} the ``non-Hermitian Berry connection'' used in other works on non-Hermitian evolution~\cite{garrison1988complex, liang2013topological}, which is defined using a non-orthonormal basis and a bi-orthogonal product.

Let us now define
\begin{equation}
  \rho_a^{(1)} \equiv \frac{\mathcal{A}_{12}^a - TC_1}
  {T\left(\lambda_1-\lambda_2\right)}.
\end{equation}
Unlike the Hermitian counterpart~\eqref{eqn:rho1}, the numerator is modified to account for the non-Hermitian contribution to the inter-band coupling.  Using this, we can integrate Eqs.~\eqref{eqn:diff_a1}--\eqref{eqn:diff_a2} to obtain
\begin{multline}
  \Delta a_1(s) = \left[ a_2 \, \rho_a^{(1)} \mathcal{U}_{12} \right]_0^s \\
  -i \int_0^s ds'\, a_1\, \rho_a^{(1)}\, \mathcal{A}_{21}^a
  - \int_0^s ds'\, a_2\, \dot\rho_a^{\,(1)}\, \mathcal{U}_{12}.
  \label{eqn:a1}
\end{multline}
Here, $\Delta a_1(s) \equiv a_1(s) - a_1(0)$.  By using the fact that $a_2(0) = 0$, we can re-incorporate the last term in Eq.~\eqref{eqn:a1} into the other two terms in a manner similar to Section~\ref{sec:perturbation}.  Define a non-Hermitian inter-band coherence factor $\rho_a$:
\begin{equation}
  \rho_a \equiv \sum_{n=1}^\infty \rho_a^{(n)}, \quad \rho_a^{(n)} \equiv 
  \frac{i\, \dot\rho_a^{\,(n-1)}}{T\left(\lambda_1-\lambda_2\right)}
  \;\;\textrm{for}\;\;n>1.
\end{equation}
For this series definition to converge, we require $|\dot\rho_a^{(n)}/\rho_a^{(n)}|\ll T|\lambda_1-\lambda_2|$.  In that case, Eq.~\eqref{eqn:a1} becomes
\begin{multline}
  \Delta a_1(s) = a_2(s) \rho_a(s) \,\mathcal{U}_{12}(s)
  -i \int_0^s ds' \, a_1 \rho_a \mathcal{A}_{21}^a.
  \label{eqn:a1new}
\end{multline}
This has almost exactly the same form as the Hermitian equation~\eqref{eqn:integ2}.  Notably, the first term on the right side of Eq.~\eqref{eqn:a1new} describes an effective hopping from $a_2$ to $a_1$, evaluated at the final time $s$.  The second term involves the same-band amplitude $a_1$, evaluated over the whole trajectory.

We can solve the integral equation with the ansatz 
\begin{equation}
  a_2(s) = - a_1(s)\, q_a(s) \, /\,  \mathcal{U}_{12}.
  \label{eqn:aansatz2}
\end{equation}
Here, $q_a(s)$ is a function to be determined, with initial value $q_a(0)=0$.  It determines the relative contributions of $|\chi_1\rangle$ and $|\chi_2\rangle$ to $|\Psi_1\rangle$; later, in Section~\ref{sec:Model}, we will see that this is precisely the quantity involved in the phenomenon of ``sudden transitions'' between non-Hermitian eigenstates.

Substituting Eq.~\eqref{eqn:aansatz2} into Eq.~\eqref{eqn:a1new} yields
\begin{equation}
  a_1(s) = a_1(0) \exp\left[ -i \int_0^s ds' \,
    \left(\mathcal{A}_{12}^a - TC_1\right) q_a(s') \right].
  \label{eqn:aansatz}
\end{equation}
Now, suppose the effective inter-band hopping---the first term on the right side of Eq.~\eqref{eqn:a1new}---is neglected.  In that case, we see from Eq.~\eqref{eqn:aansatz} that
\begin{equation}
  q_a \rightarrow \frac{\rho_a \mathcal{A}_{21}^a}{\mathcal{A}_{12}^a - TC_1}.
  \label{eqn:qlim}
\end{equation}
We can regard the deviation of $q_a$ from \eqref{eqn:qlim} as being ``generated'' by the effective inter-band hopping.  We call Eq.~\eqref{eqn:qlim} the ``leading-order approximation'' for the non-Hermitian evolution, in which the effective inter-band hopping is neglected.  (Note that this approximation violates the initial value condition for $q_a$.)  

We can go beyond the leading-order approximation by converting Eqs.~(\ref{eqn:a1new})--\eqref{eqn:aansatz} into differential form:
\begin{equation}
  -i\frac{dq_a}{ds} = - \mathcal{A}_{21}^a + 
 T\big(\lambda_1-\lambda_2\big)q_a + 
 \left(\mathcal{A}_{12}^a - TC_1\right)q_a^2.
  \label{eqn:qa}
\end{equation}
Appendix~\ref{Solution.apx} describes a systematic semi-analytic procedure for solving this first-order nonlinear differential equation.  We find that it is convenient to make a ``sub-leading-order approximation'' that involves breaking $q_a$ into two terms:
\begin{equation}
  q_a \approx \bar{q}_a
  +\frac{\tilde{q}_a(0)\,e^{i(\Omega_1-\Omega_2)}}{1-\tilde{q}_a(0)\,
    \int_0^s ds'\;e^{i(\Omega_1-\Omega_2)}},
  \label{eqn:diff1a}
\end{equation}
where the first term is a slowly-varying part and the second term is a rapidly oscillating and decaying part.  Alternatively, Eq.~\eqref{eqn:qa} can also be solved fully numerically.

Next, we consider the second component of Eq.~(\ref{eqn:psi_0}).  We will handle this using the basis $\{|\eta_2(s)\rangle,|\xi_2(s)\rangle\}$.  Let $|\Psi_2(s)\rangle$ denote this part of the state vector:
\begin{equation}
  |\Psi_2(s)\rangle=b_1(s)e^{-i\Omega_2}|\xi_2(s)\rangle
  + b_2(s)e^{-i\Omega_1}|\eta_2(s)\rangle.
\label{eqn:psi_b}
\end{equation}
Substitution into the time-dependent Schr\"odinger equation, and left-multiplying by $\langle\eta_2(s)|$ and $\langle\xi_2(s)|$, yields
\begin{align}
  \begin{aligned}
    \dot{b}_1 & = ib_2\, \mathcal{U}_{21}
    \Big( \mathcal{A}_{21}^b - TC_2\Big) \\ 
    \dot{b}_2 &= ib_1 \, \mathcal{U}_{12} \mathcal{A}_{12}^b,
  \end{aligned}
  \label{eqn:diff_b1}
\end{align}
where
\begin{equation}
  \mathcal{A}_{12}^b \equiv i\langle\eta_2|\dot{\xi}_2\rangle,\;\;
  \mathcal{A}_{21}^b \equiv i\langle\xi_2|\dot{\eta}_2\rangle.
\end{equation}
Repeating the preceding procedure for the $b_2$ amplitude (which characterizes the most amplifying state) gives
\begin{multline}
  \Delta b_2(s) = b_1 \rho_b \,\mathcal{U}_{12}
  -i\int_0^s ds' \, b_2 \rho_b \Big(\mathcal{A}_{21}^b - TC_2\Big),
  \label{eqn:diff_b2}
\end{multline}
where $\Delta b_2(s) \equiv b_2(s) - b_2(0)$, and
\begin{multline}
    \rho_b \equiv \sum_{n=1}^\infty \rho_b^{(n)}, \quad\;
    \rho_b^{(1)} \equiv \frac{\mathcal{A}_{12}^b}{T\left(\lambda_1-\lambda_2\right)}, \\
    \rho_b^{(n)} \equiv \frac{i\dot\rho_b^{\,(n-1)}}{T\left(\lambda_1-\lambda_2\right)}
    \;\;\textrm{for}\;\;n > 1 \;.
\end{multline}    
Note that the inter-band coherence term is defined differently for the $b$ amplitudes than for the $a$ amplitudes.  The second (``intra-band'') term on the right side of Eq.~\eqref{eqn:diff_b2} also has a slightly different form from Eq.~\eqref{eqn:a1new}.  In order for the series definition of $\rho_b^{(n)}$ to converge, we must have $|\dot\rho_b^{(n)}/\rho_b^{(n)}|\ll T|\lambda_1-\lambda_2|$.

We now take the ansatz
\begin{equation}
  b_1(s) = - b_2(s)\, q_b(s) \,/\, \mathcal{U}_{12},
  \label{bansatz}
\end{equation}
where $q_b(s)$ is a function to be determined.  Substituting this into Eq.~\eqref{eqn:diff_b2} gives
\begin{equation}
  b_2(s) = b_2(0) \exp\left[-i \int_0^s ds'\; \mathcal{A}^b_{12}(s') \, q_b(s')\right].
  \label{eqn:bansatz}
\end{equation}
If the effective inter-band hopping---described by the first term on the right of Eq.~\eqref{eqn:diff_b2}---is negligible, then
\begin{equation}
  q_b \rightarrow \frac{\rho_b \left(\mathcal{A}_{21}^b - TC_2\right)}{\mathcal{A}_{12}^b}.
  \label{eqn:qblim}
\end{equation}
As before, we call \eqref{eqn:qblim} the ``leading-order approximation'', equivalent to neglecting the inter-band couplings.

To go beyond the leading-order approximation, we convert Eqs.~\eqref{eqn:diff_b2}--\eqref{bansatz} into differential form:
\begin{equation}
  -i \frac{dq_b}{ds} 
  = -\left(\mathcal{A}_{21}^b - TC_2\right) + 
  T\big(\lambda_1-\lambda_2\big)q_b +
  \mathcal{A}_{12}^b \, q_b^2\;.
  \label{eqn:qb}
\end{equation}
Similar to Eq.~\eqref{eqn:qa}, this can be solved numerically or semi-analytically.  In the semi-analytic solution, a sub-leading approximation can be defined by breaking the solution into slowly and rapidly-varying parts:
\begin{equation}
  q_b \,\approx\, \bar{q}_b
  + \frac{\tilde{q}_b(0)\,e^{i(\Omega_1-\Omega_2)}}{1-\tilde{q}_b(0)\,\int_0^s ds'\;e^{i(\Omega_1-\Omega_2)}}\;.
  \label{eqn:diff1b}
\end{equation}

In summary, we find that the time-dependent system state can be written as two parts,
\begin{equation}
  |\Psi(s)\rangle = |\Psi_a(s)\rangle +  |\Psi_b(s)\rangle,
  \label{eqn:result0}
\end{equation}
where
\begin{align}
  |\Psi_a\rangle & = a_1(0)\, S_a\, e^{-i\Omega_1} \Big( |\chi_1\rangle - q_a\, |\chi_2\rangle\Big) \label{eqn:result1a} \\
  |\Psi_b\rangle & = b_2(0)\, S_b\, e^{-i\Omega_1}\,
  \Big( |\eta_2\,\rangle - q_b\, |\xi_2\,\rangle \Big),
  \label{eqn:result1b}
\end{align}
with $S_{a,b}$ and $q_{a,b}$ defined as follows:
\begin{align}
    S_a(s) &= \exp
    \left[ -i \int_0^s ds' \, \left(\mathcal{A}_{12}^a - TC_1\right) q_a(s') \right]
    \label{result_sa} \\
    S_b(s) &= \exp
    \left[-i \int_0^s ds'\; \mathcal{A}^b_{12}(s') \, q_b(s')\right].
    \label{result_sb}
\end{align}
In Eqs.~\eqref{eqn:result1a}--\eqref{eqn:result1b}, the leading scale factor of $\exp(-i\Omega_1)$ corresponds to the fastest amplification rate present in the system.  The ``direction'' of the state vector is mainly determined by the magnitudes of $q_a$ and $q_b$.  In the leading-order approximation, these are given by Eqs.~\eqref{eqn:qlim} and \eqref{eqn:qblim} respectively.  In the sub-leading-order approximation, we replace these phase shifts with Eqs.~\eqref{eqn:diff1a} and \eqref{eqn:diff1b}, as described in Appendix~\ref{Solution.apx}.  For a higher-order approximation, we solve for $q_{a}$ and $q_b$ using the full differential equations \eqref{eqn:qa} and \eqref{eqn:qb}.

\subsection{$n\times n$ non-Hermitian Hamiltonian}

For a general $n\times n$ non-Hermitian Hamiltonian, we first use a Schur decomposition to find an orthonormal basis $\{|\chi_j\rangle\}$, with which the Hamiltonian is written in the form of Eq.~\eqref{eqn:SchurH}.   We then use a succession of different Schur decompositions, as described in Section~\ref{sec:Schur}, to construct a basis $\{|\chi_1\rangle,|\eta_2\rangle,\dots,|\eta_n\rangle\}$ that is non-orthogonal but complete.  We use this basis to decompose the initial state vector, and follow the dynamics of each component separately.  Let the initial state be
\begin{equation}
  |\Psi(0)\rangle = c_1(0)|\chi_1(0)\rangle + \sum_{j=2}^{n} c_j(0) |\eta_j(0)\rangle.
\end{equation}
We use the orthonormal basis $\{|\chi_j\rangle\}$ to handle the first component:
\begin{equation}
  |\Psi_1(s)\rangle = \sum_{j=1}^{n}a_j(s) e^{-i\Omega_j} |\chi_j(s)\rangle,
\end{equation}
where $a_1(0)=c_1(0)$ and $a_j(0)=0$ for $j>1$.  Substituting this into the Schr\"odinger equation, and left-multiplying by $\{\langle\chi_j|\}$, yields
\begin{equation}
  \dot{a}_j = i\sum_{k\ne j}^{}a_k\, \mathcal{U}_{jk}\mathcal{C}_{jk}^a, \;\;
  \mathrm{where}\;\;
  \mathcal{C}_{jk}^a \equiv \mathcal{A}_{jk}^a - TC_{jk}.
  \label{eqn:diffN}
\end{equation} 
Note that $C_{jk}= 0$ for $j\geqslant k$.  The integral form is
\begin{equation}
  \Delta a_1(s) = \sum_{j=2}^{n} a_j\rho_{1j}^a\,\mathcal{U}_{1j}
  -i \int_0^s ds' \, a_1(s') \sum_{j=2}^{n}\,\rho_{1j}^a \mathcal{C}_{j1}^a,
  \label{eqn:integN}
\end{equation}
where
\begin{equation}
  \rho_{1j}^a \equiv \sum_{n=1}^\infty \rho_{1j}^{a(n)}, \quad
  \rho_{1j}^{a(1)} \equiv \frac{\mathcal{C}_{1j}^a}{T\left(\lambda_1-\lambda_j\right)},
\end{equation}
and, for $n > 1$,
\begin{equation}
  \rho_{1j}^{a(n)} \equiv \frac{i}{T\left(\lambda_1-\lambda_j\right)} \left[ \dot\rho_{1j}^{\,a(n-1)} 
    + i\,\sum\limits_{\ell} \rho_{1\ell}^{a(n-1)} \mathcal{C}_{\ell j}^a \right].
\end{equation}
To solve this, we take the ansatz
\begin{equation}
  a_j(s) = - a_1(s)\, q_j^a(s) \, /\,  \mathcal{U}_{1j}\;\;\;\textrm{for}\;\;j > 1.
    \label{cansatz}
\end{equation}
The leading approximation is defined by
\begin{equation}
  q_j^a \rightarrow \frac{\rho_{1j}^a\mathcal{C}_{j1}^a}{\mathcal{C}_{1j}^a},
\end{equation}
and an improved approximation is obtained by substituting Eq.~\eqref{cansatz} into Eq.~(\ref{eqn:diffN}) to obtain
\begin{equation}
  i\dot q_j^a = -C_{j1}+ T\big(\lambda_1-\lambda_j\big)q_j^a +
  \sum\limits_{\ell=2}^{n}\,\left(C_{j\ell}^a+q_j^a\,C_{1\ell}^a\right)\,q_{\ell}^a.
  \label{eqn:iterativeN2}
\end{equation}
When $\left| \frac{d}{ds} \left( q_j^a\,C_{1j}^a \right) / \left( q_j^a\,C_{1j}^a \right) \right|\ll T|\lambda_1-\lambda_j|$, Eq.~\eqref{eqn:iterativeN2} can be solved iteratively to obtain $q_j^a(s)$.  Similar to the $2\times 2$ case, this semi-analytic solution can be decomposed into a slowly-varying part and rapidly oscillating and decaying part determined by the initial conditions.  The remaining components are handled in a manner similar to Section~\ref{2by2}.

\section{Numerical results}
\label{sec:Model}

To test the method derived in the previous section, we consider the specific $2\times2$ non-Hermitian Hamiltonian
\begin{equation}
  H(t)=\begin{pmatrix}
  \omega(t)+i\nu(t) & c\\
  c & -\omega(t)-i\nu(t)\\
  \end{pmatrix},
\label{eqn:forwardH0}
\end{equation}
where $\omega(t)$ and $\nu(t)$ are real time-dependent parameters, and $c$ is a constant.  The eigenvalues are
\begin{equation}
  \lambda=\pm\sqrt{c^2+\left(\omega+i\nu\right)^2}.
\end{equation}
At the exceptional points $\omega+i\nu=\pm ic$, the eigenvalues and their associated eigenvectors coalesce.  The corresponding eigenstates have the form
\begin{equation}
  |\psi(t)\rangle \propto \begin{pmatrix}1 \\ z \end{pmatrix},
  \label{eqn:psit}
\end{equation}
where $z\equiv x+i\,y$ is the complex ratio of the components.

\begin{figure}
  \centering       
  \includegraphics[width=\columnwidth]{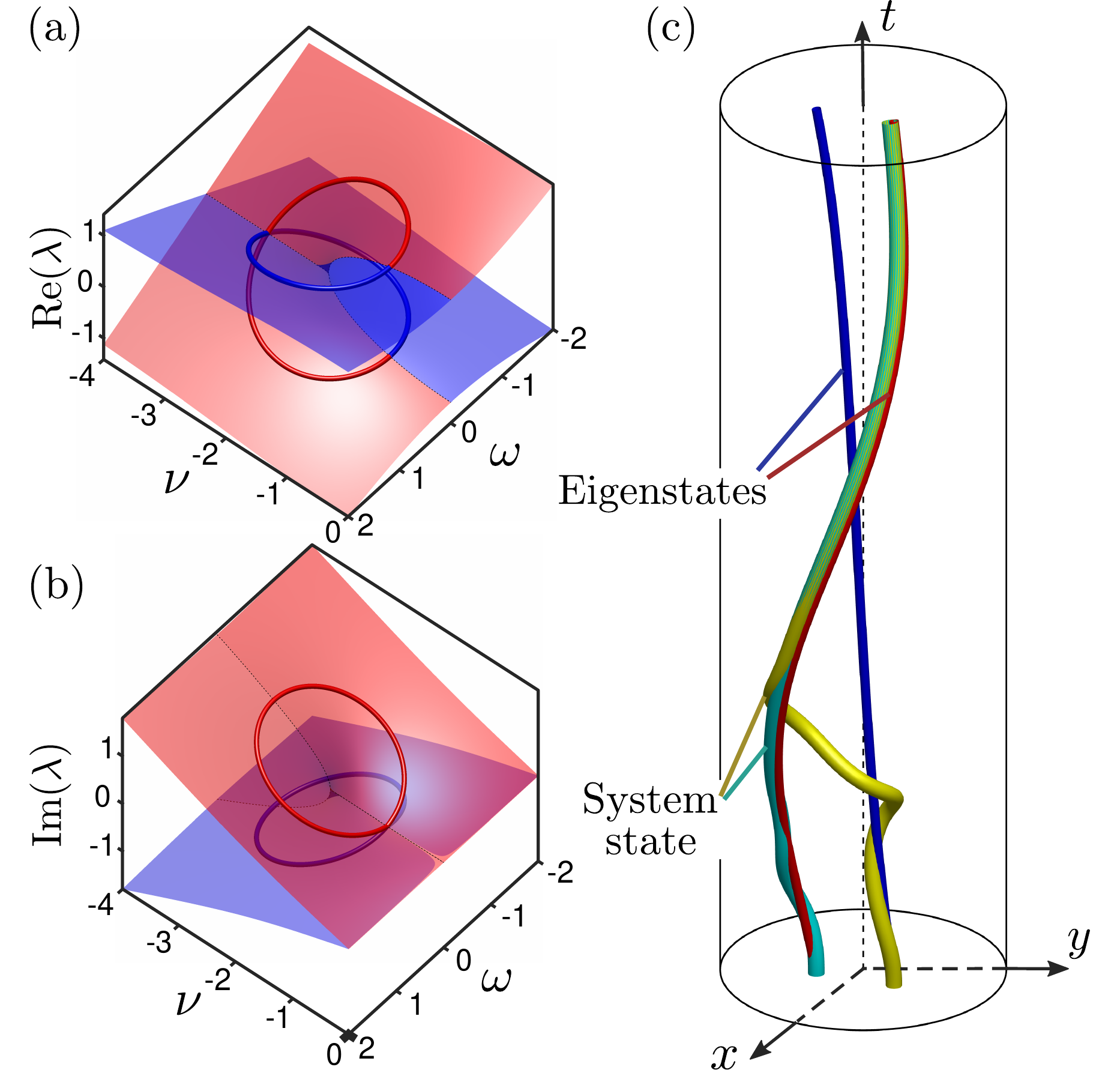}
  \caption{(a)--(b) Complex eigenvalue spectrum of the non-Hermitian Hamiltonian $H$, versus the parameters $\omega$ and $\nu$, for constant $c = 2$.  Here, we show the region of parameter space near the exceptional point at $\omega = 0, \; \nu = -2$.  A trajectory loop $\omega+i\nu = e^{2i\pi t/T}-ic$, encircling the exceptional point, is indicated.  Amplifying (decaying) states are indicated in red (blue).  (c) Instantaneous eigenstates and time-evolving system states, characterized by the complex number $z = x +iy$ defined in Eq.~\eqref{eqn:psit}, versus time $t$.  The red and blue lines respectively indicate amplifying and decaying instantaneous eigenstates of $H(t)$, as the Hamiltonian slowly undergoes the trajectory indicated in (a)--(b), with time scale $T=10$.  The cyan and yellow lines represent the evolving system states, with initial conditions set to either the amplifying (cyan) or decaying (yellow) initial instantaneous eigenstate.}
  \label{fig:trajectory}
\end{figure}

\begin{figure}
  \centering
  \includegraphics[width=\linewidth]{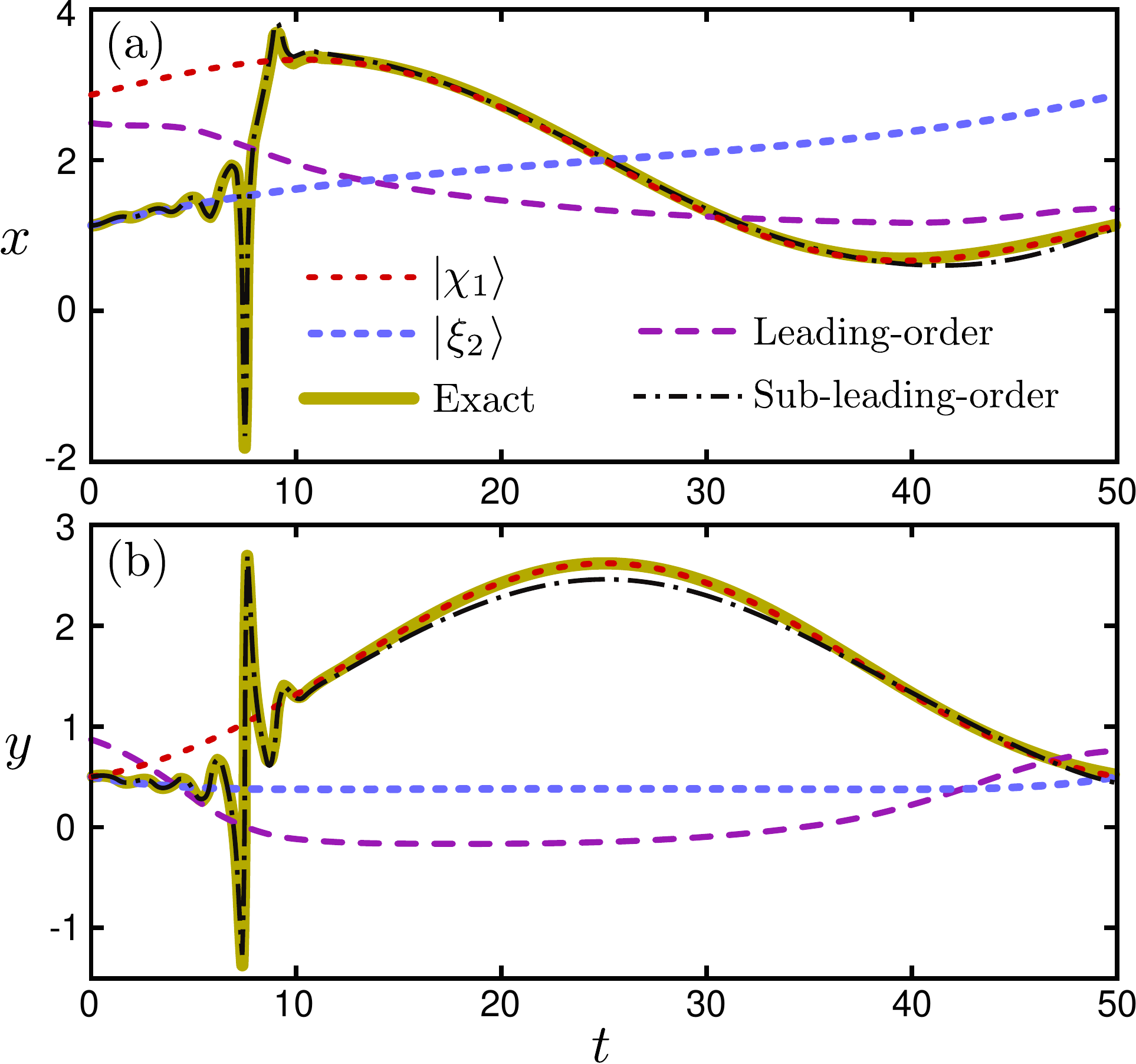}
  \caption{Time evolution of the system state as the Hamiltonian cycles once around the exceptional point, with $c = 2$, $T = 50$, and $\omega+i\nu\equiv e^{2i\pi t/T}-ic$.  The system state is initialized in the decaying instantaneous eigenstate at $t = 0$.  The plotted quantities (a) $x$ and (b) $y$ characterize the system state up to a scale factor, as defined in Eq.~\eqref{eqn:psit}.  Direct numerical integration shows that the time-evolving system state undergoes a sudden transition to the amplifying eigenstate at $t \approx 7.5$.  This transition is captured by the sub-leading-order approximation developed in the text.}
  \label{fig:diabatic_state}
\end{figure}

Figure~\ref{fig:trajectory}(a)--(b) shows the complex eigenvalue spectrum of $H$, as a function of $\omega$ and $\nu$.  Red and blue colors indicate eigenstates that are amplifying ($\mathrm{Im}\lambda > 0$) and decaying ($\mathrm{Im}\lambda < 0$) respectively.  One particular parametric trajectory, $\omega+i\nu = e^{2i\pi t/T}-ic$, is also shown; note that this trajectory encircles an exceptional point.  Fig.~\ref{fig:trajectory}(c) plots $x$ and $y$---the real and imaginary parts of the eigenvector component $z$ defined in Eq.~\eqref{eqn:psit}---versus the time $t$, as the Hamiltonian goes through this parametric trajectory.  Encircling the exceptional point once leads to an interchange of the two instantaneous eigenstates~\cite{Heiss1999,Dembowski2001}.

The behavior of the system state under actual time evolution, however, is more subtle. The yellow curve in Fig.~\ref{fig:trajectory}(c) shows the dynamical state, computed by integrating the Schr\"odinger equation numerically using the split-step method (the results of which can be considered ``exact'', apart from the usual small discretization errors from numerical integration). It is observed that if the initial system state is in either of the initial instantaneous eigenstates, then for sufficiently slow Hamiltonian variation and short elapsed times, the state clings to the instantaneous eigenstate, similar to the adiabatic limit of Hermitian dynamics.  For longer times, however, the system can undergo a sudden transition from a decaying eigenstate to an amplifying eigenstate. This ``breakdown of adiabaticity'' has been extensively commented upon in previous works~\cite{moiseyev2011non, uzdin2011observability, berry2011slow, heiss2016mathematical, doppler2016dynamically, xu2016topological, gong2016aharonov}, and may be technologically useful as a means of realizing high-efficiency nonlinear optical isolators \cite{choi2017extremely}.

Figure~\ref{fig:diabatic_state} compares the exact results to the results from the evolution equations derived in Section~\ref{sec:nonhermitian}.  The latter are calculated using the non-Abelian connections, eigenvalues, and Schur components derived from the Hamiltonian (and its Schur decompositions) at each instant along the trajectory.

According to Eqs.~\eqref{eqn:result1a}--\eqref{eqn:result1b}, the transitions are governed by the quantities $q_a$ and $q_b$, which describe the relative proportions of the eigenstate contributions to the state vector.  We observe that the leading-order approximation (from neglecting the effective inter-band hopping) fails to match the exact results.  However, the sub-leading approximation (which employs explicit but approximate expressions for $q_a$ and $q_b$) agrees well with the exact results.  In particular, it accurately describes the sudden transition from clinging to the decaying eigenstate to clinging to the amplifying eigenstate.  In the sub-leading-order approximation, the relative errors of $|q_a|$ and $|q_b|$ are less than $0.3\%$ in this numerical example.

\begin{figure}
  \centering
  \includegraphics[width=\linewidth]{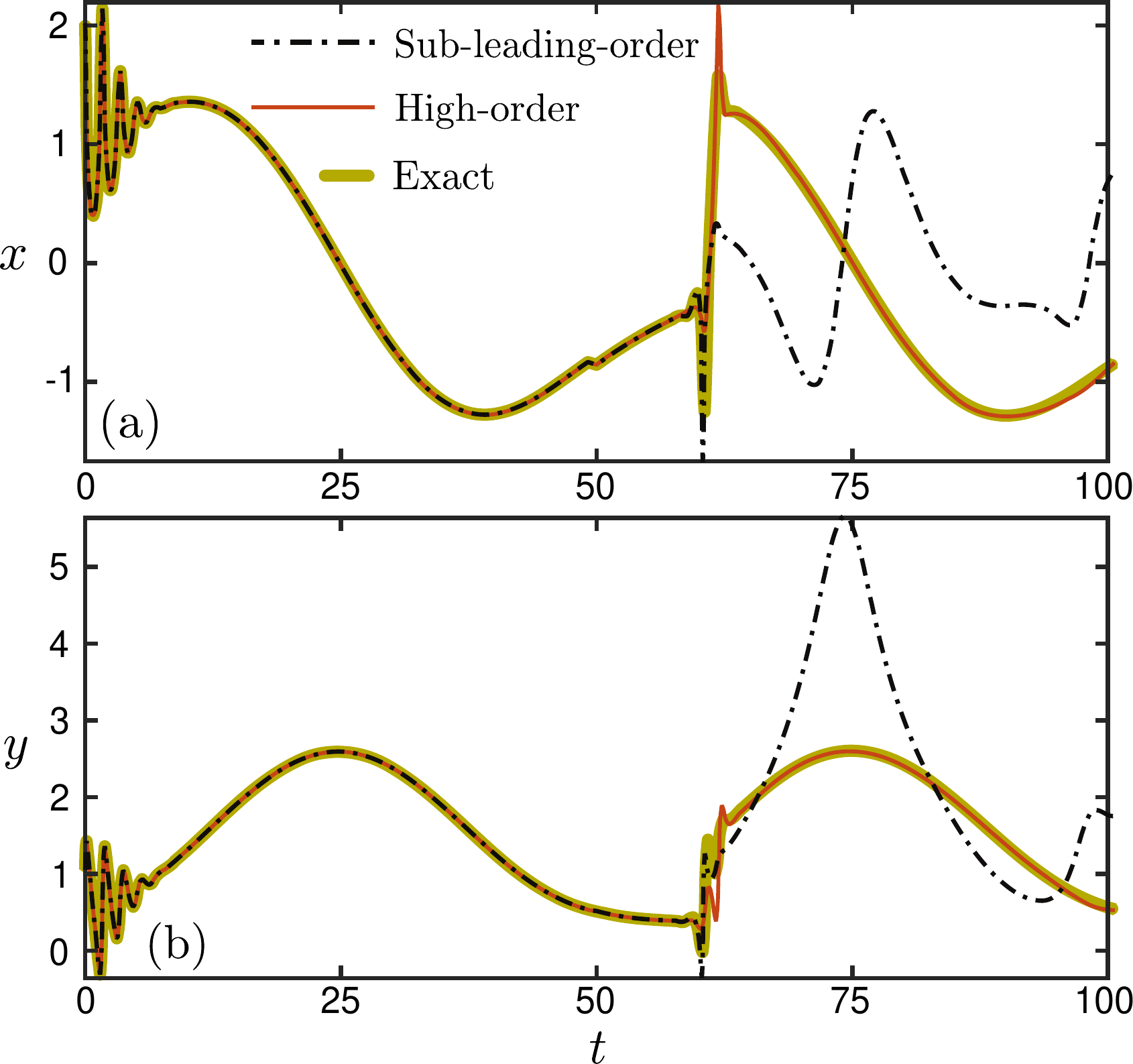}
  \caption{Time evolution of the system state as the Hamiltonian cycles twice around the exceptional point, with $c = 2$, $T = 50$, and $\omega+i\nu\equiv e^{2i\pi t/T}-ic$, with the system state initialized in the amplifying instantaneous eigenstate at $t = 0$.  The plotted quantities (a) $x$ and (b) $y$ characterize the system state up to a scale factor.  Results are shown for exact numerical integration of the Schr\"odinger equation, the sub-leading-order approximation, and a high-order approximation which solves $q_a$ and $q_b$ via the method described in Appendix~\ref{Solution.apx}.}
  \label{fig:diabatic_state2}
\end{figure}

The sub-leading-order approximation can also break down for sufficiently long elapsed time.  Fig.~\ref{fig:diabatic_state2} shows the evolution for the same system, with the state initialized to the \textit{amplifying} initial instantaneous eigenstate.  At the end of the first cycle around the exceptional point, the state is clinging to the decaying eigenstate (due to eigenstate exchange).  Shortly into the second cycle, it undergoes a sudden transition to the amplifying eigenstate.  The sub-leading-order approximation gives the correct transition time, but fails to accurately describe the subsequent evolution.  This failure is due to the accumulation of approximation errors over long elapased times, which can be reduced by either using a slower evolution protocol, or by introducing a higher-order approximation as described in Appendix~\ref{Solution.apx}.

\begin{figure}
  \centering       
  \includegraphics[width=\columnwidth]{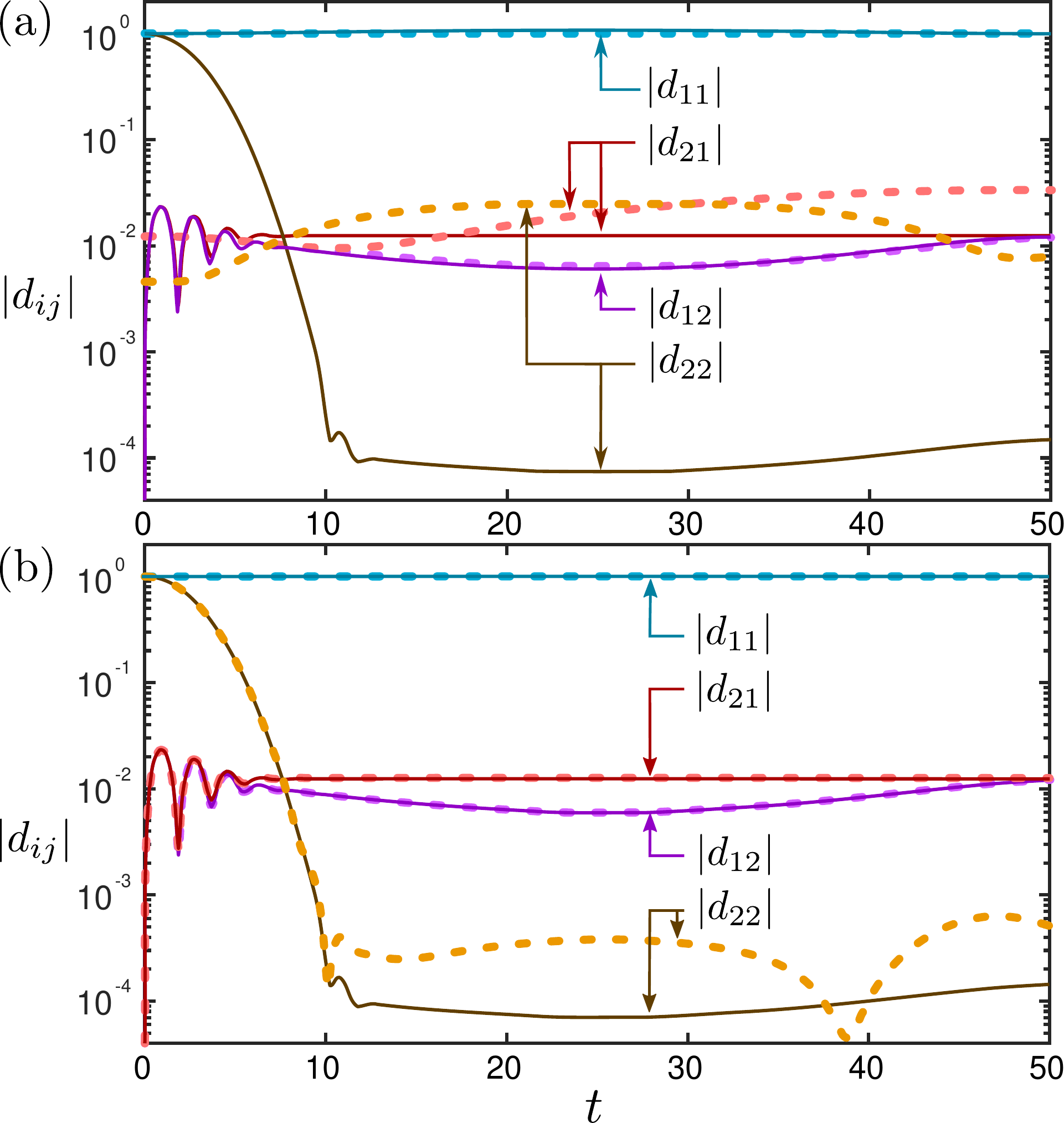}
  \caption{Magnitude of the adiabatic multipliers, $|d_{ij}|$, versus time $t$ for the $2\times2$ model with $\omega+i\nu\equiv e^{2i\pi t/T}-ic$, $c=2$, and $T=50$.  The solid lines are obtained by direct numerical integration, while the dashes are calculated via Eqs.~\eqref{eqn:d11}--\eqref{eqn:d22}.  In (a), the leading-order approximation is used, and in (b) the sub-leading-order approximation is used.  The exponential variation in $|d_{22}|$ corresponds to the sudden transition which occurs at $t \approx 7.5$ when the system is initialized in the decaying eigenstate, as shown in Fig.~\eqref{fig:diabatic_state}.}
    \label{fig:diabatic_multiplier1}
\end{figure}

These results can also be analyzed through the ``adiabatic multiplier'' concept previously used by Berry and Uzdin to quantify sudden transitions in non-Hermitian dynamics~\cite{berry2011slow,dingle1973asymptotic}.  Let the initial system state be
\begin{equation}
  |\Psi(0)\rangle = c_1(0)|\chi_1(0)\rangle + c_2(0) |\xi_2(0)\rangle,
\end{equation}
where $|\chi_1(0)\rangle$ and $|\xi_2(0)\rangle$ are the instantaneous eigenstates of $H(0)$ (see Section~\ref{sec:Schur}), which is assumed not to be at an exceptional point; $c_1(0)$ and $c_2(0)$ are the corresponding complex amplitudes.  The state at subsequent times can be written in the form
\begin{multline}
  |\Psi(s)\rangle = e^{-i\Omega_1(s)}\bigg[\Big(c_1(0)d_{11}(s)+c_2(0)d_{21}(s)\Big)\,|\chi_1(s)\rangle\\
   + \Big(c_1(0)d_{12}(s)+c_2(0)d_{22}(s)\Big)\,|\xi_2(s)\rangle\bigg],
     \label{eqn:eigendecomposition}
\end{multline}
where $\Omega_1(s) = T \int_0^s ds' \lambda_1(s')$ is the dynamical factor corresponding to the most amplifying eigenstate.  The $d_{ij}(s)$ coefficients are the adiabatic multipliers.  

By direct substitution, it can be shown that the adiabatic multipliers can be written in terms of the quantities appearing in our Eqs.~\eqref{eqn:result0}--\eqref{result_sb} (i.e., Schur coefficients, inter-band coherence factors, etc.):
\begin{align}
  d_{11} & = S_a \left(1-\frac{q_a\,C_1}{\Delta\lambda}\right) \label{eqn:d11}\\
  d_{12} & = -\frac{S_a\,q_a}{\langle\chi_2|\xi_2\rangle}\\
  d_{21} & = \frac{d_{11}}{\eta_{21}^0}
  - \frac{\Delta \lambda^0\, S_b}{C_2(0) \langle\eta_2|\chi_1\rangle} \\
  d_{22} & = \frac{d_{12}}{\eta_{21}^0} +\frac{\Delta\lambda^0\, S_b}{C_2(0)}
  \Big( q_b-\frac{C_2}{\Delta\lambda}\Big), \label{eqn:d22}
\end{align}
where $\eta_{21}^0\equiv \langle\xi_2(0)|\chi_1(0)\rangle$, $\Delta\lambda^0 \equiv \lambda_1(0)-\lambda_2(0)$, and $\Delta\lambda \equiv \lambda_1(s)-\lambda_2(s)$.

If the system starts in the amplifying state [$c_2(0) = 0$], sudden transitions occur when $d_{11}$ and/or $d_{12}$ undergo exponential variations; if the system starts in the decaying state [$c_1(0) = 0$], sudden transitions occur with exponential variations in $d_{21}$ and/or $d_{22}$.

Figure~\ref{fig:diabatic_multiplier1} plots the magnitude of the adiabatic multipliers for the same $2\times2$ non-Hermitian Hamiltonian as before.  In Fig.~\ref{fig:diabatic_multiplier1}(a), the leading-order approximation is used.  At long times, the adiabatic multipliers $|d_{11}|$ and $|d_{12}|$ produced by the leading-order approximation agree with the exact values (solid lines), whereas $|d_{21}|$ and $|d_{22}|$ do not match at all.  This is consistent with our earlier findings that the leading-order approximation does not give a good description of the time evolution.  In Fig.~\ref{fig:diabatic_multiplier1}(b), the sub-leading-order approximation is used.  Now we obtain excellent agreement with the exact results (solid lines), with the only notable deviations occurring at very small values of $|d_{22}|$.  (These small deviations can be further reduced if higher-order approximations are taken.)  The observed exponential decrease in $|d_{22}|$, from unity to nearly zero, corresponds to the sudden transition from the decaying to the amplifying state in Fig.~\ref{fig:diabatic_state}.

\section{Conclusions}
\label{sec:conclu}

We have developed a theoretical framework for describing the time evolution of a general non-Hermitian system.  We obtained explicit closed-form expressions for the quantum amplitudes,  involving the instantaneous complex energies and inter-band Berry connections.  In particular, the Berry connections are defined with regular inner products, using orthonormal basis vectors produced by Schur decompositions of the non-Hermitian Hamiltonian, rather than the bi-orthogonal products employed in previous works on non-Hermitian evolution~\cite{garrison1988complex, liang2013topological}.  Unlike previous studies that used such generalized Berry connection to describe non-Hermitian dynamics~\cite{mehri2008geometric, gong2010geometric, gong2013time}, our theory is not restricted to the special case of cyclic Hamiltonians and dynamical states that return to themselves after one cycle.

We have shown numerically that our theory accurately describes the phenomenon of ``sudden transitions'' in non-Hermitian dynamics, where the system state jumps from one non-Hermitian eigenstate to another~\cite{moiseyev2011non, uzdin2011observability, berry2011slow, gong2016aharonov, heiss2016mathematical, doppler2016dynamically, xu2016topological}. This phenomenon has recently been shown to be useful for realizing efficient optical isolators~\cite{choi2017extremely}.  In our theory, the key role in these transitions is played by the complex functions $q_{a}$ and $q_b$, which are affected precisely by those terms in the quantum amplitude equations that describe effective inter-band hoppings.  In future work, it would be important to examine these functions in greater detail, and try to develop a better physical understanding of them. It would also be interesting to use our theory to analyze sudden transitions in more complicated non-Hermitian systems, such as periodically-driven non-Hermitian Hamiltonians  \cite{aharonov1987phase,garrison1988complex} or Hamiltonians with high-order exceptional points \cite{chen2017exceptional, hodaei2017enhanced, ding2016emergence, jing2017high}.

\section{Acknowledgements}

We thank Longwen Zhou, Qinghai Wang and Jiangbin Gong for useful discussions and comments.  This work was supported by the Singapore MOE Academic Research Fund Tier 2 Grant No.~MOE2015-T2-2-008, and the Singapore MOE Academic Research Fund Tier 3 Grant No.~MOE2016-T3-1-006.

\appendix

\section{Unitary transformation of an upper triangular matrix}
\label{Utransformation.apx}

Consider an upper triangular matrix $A$ of the form
\begin{equation}
  A_{i,j}=\left\{
  \begin{array}{rcl}
    C_{i,j} & \text{for} & i<j\\
    \lambda_j & \text{for} & i=j\\
    0 & \text{for} & i>j
\end{array}\right..
\end{equation} 
We can swap any of the two neighboring diagonal entities, i.e. $\lambda_{j-1}$ and $\lambda_{j}$, through a unitary transformation
\begin{equation}
  U_j=\begin{pmatrix}
  \mathbb{I}^{j-2} &  &  \\
  & W^{(j)} & \\
  & & \mathbb{I}^{n-j}
  \end{pmatrix},
\end{equation}
where $\mathbb{I}^{n}$ denotes a $n\times n$ identity matrix,
\begin{equation}
  W^{(j)} = \frac{1}{\sqrt{1+|z_j|^2}}
  \begin{pmatrix} 
    1 & -z_j^* \\
    z_j & 1
  \end{pmatrix}
\end{equation}
and 
\begin{equation}
  z_j = \frac{\lambda_j-\lambda_{j-1}}{C_{j-1,j}} \;,
\end{equation}
so that the sequence of the diagonal entries become $\lambda_1,\cdots,\lambda_{j-2}, \lambda_j, \lambda_{j-1},\lambda_{j+1},\cdots$.  In the special case when $C_{j-1,j}=0$, we use $[0,1;1,0]$ in place of $W^{(j)}$.  

For each $j$, we can always use a succession of $j-1$ such transformations to bring the eigenvalue $\lambda_j$ to the front, so that the sequence is $\lambda_j, \lambda_1, \cdots,\lambda_{j-1},\lambda_{j+1}, \cdots$.

\section{Completeness of the basis set}
\label{completeness.apx}

The set $\{|\chi_1\rangle,|\eta_2\rangle,\cdots,|\eta_n\rangle\}$ introduced in Section~\ref{sec:Schur} is complete so long as the eigenvector $|\chi_1\rangle$ is non-orthogonal to every other other eigenvector $|\xi_{j}\rangle$ ($j = 2, \cdots, n$).  To prove this, we start from the Schur decomposition theorem, which states that $\{\,|\chi_1\rangle, |\chi_2\rangle, \cdots, |\chi_n\rangle\, \}$ is complete and orthogonal.  Using the unitary transformation scheme introduced in Appendix~\ref{Utransformation.apx}, we can use a succession of $j-2$ unitary transformations to re-arrange the diagonal entries of the Schur form $A$ to the desired order, i.e. $\lambda_1, \lambda_j, \lambda_2, \cdots,\lambda_{j-1},\lambda_{j+1}, \cdots$.  In the basis defined by this new Schur decomposition, the first basis vector is still $|\chi_1\rangle$.  The second basis vector is a superposition of $\{|\chi_2\rangle,|\chi_3\rangle,\cdots,|\chi_j\rangle\}$.

We can further swap the first two diagonal entries, i.e. $\lambda_1$ and $\lambda_j$, through a unitary transformation to bring $\lambda_j$ to the front.  In the notation introduced in Section~\ref{sec:Schur}, after the transformation the first basis vector $|\xi_j\rangle$ is an eigenvector with eigenvalue $\lambda_j$, while the second basis vector $|\eta_j\rangle$ is ``associated'' with $\lambda_1$.  From our assumption that $|\chi_1\rangle$ is non-orthogonal to each of the other eigenvectors, we can show that $|\eta_j\rangle$ is a superposition of the vectors $\{|\chi_1\rangle,|\chi_2\rangle,\cdots,|\chi_j\rangle\}$, and importantly the composition of $|\chi_j\rangle$ is nonzero.  In this way, we see that the dimension of the subspace spanned by $\{|\chi_1\rangle,|\eta_2\rangle,\cdots,|\eta_j\rangle\}$ increases monotonically with $j$.  This shows that $\{|\chi_1\rangle,|\eta_2\rangle,\cdots,|\eta_n\rangle\}$ forms a complete basis for the space spanned by $\{\,|\chi_1\rangle, |\chi_2\rangle, \cdots, |\chi_n\rangle\, \}$.

\section{Iterative solution method}
\label{Solution.apx}

As noted in the main text, the evolution of a non-Hermitian system can be characterized using a set of functions, denoted by $q_n$, that describe the relative contributions of the different Schur basis vectors.  In the $2\times2$ case, we have the $q_a$ function that satisfies Eq.~\eqref{eqn:qa}, and the $q_b$ function that satisfies Eq.~\eqref{eqn:qb}.  These first-order nonlinear differential equations have the form
\begin{equation}
-i\,\dot q=-A+B\,q+C\,q^2.
\label{eqn:CDE1}
\end{equation}
To solve this, let us define
\begin{align}
  q_1 & \equiv A/B\\
  A_1 & \equiv-i\,\dot{q}_1-C\,q_1^2\\
  B_1 & \equiv B+2\,C\,q_1.
\end{align}
Then Eq.~(\ref{eqn:CDE1}) becomes
\begin{equation}
-i\,\dot{q'}=-A_1+B_1\,q'+C\,q'^2\;,
\end{equation}
where $q'=q-q_1$.  We can now iteratively define 
\begin{align}
  q_{j+1} & \equiv A_j/B_j\\
  A_{j+1} & \equiv-i\,\dot{q}_{j+1}-C\,q_{j+1}^2\\
  B_{j+1} & \equiv B_j+2\,C\,q_{j+1}
\end{align}
and hence recast Eq.~(\ref{eqn:CDE1}) as
\begin{equation}
-i\,\dot{\tilde q} = \tilde B\,\tilde q+C\,{\tilde q}^2,
\label{eqn:CDE2}
\end{equation}
where $\tilde B = B+2\,C\,\sum_{j=1}^{\infty} q_j$.  Eq.~(\ref{eqn:CDE2}) can be solved exactly, and the solution is 
\begin{equation}
  \tilde q = \frac{\tilde q(0) \exp\left[i\int_0^s\tilde B ds'\right]}
           {1-\tilde q(0)\int_0^s iC\exp\left[i\int_0^{s'}\tilde B ds''\right]ds'}\;,
\end{equation}
where $\tilde q(0)$ is determined by the initial value condition for $q$.  Thereby, the solution to Eq.~(\ref{eqn:CDE1}) is given by $q=\bar q+\tilde q$.  The first part $\bar q\equiv\sum_{j=1}^{\infty} q_j$, is a slowly-varying series; the second part -- $\tilde q$, is a rapidly oscillating and decaying function.

We note that $AC/B^2\propto T^{-1}$.  In our problem, the variable $T$ characterizes the time scale for the Hamiltonian's trajectory within some parameter space.  For a sufficiently slowly varying Hamiltonian, $q_j\propto T^{-j}$ obeys power-law decay and goes rapidly to zero.  The series $q_j$ thus constitute a hierarchy of solutions that converges for large $T$.  If the condition $|\dot{q}_1/q_1|\ll T|\lambda_2-\lambda_1|$ is satisfied, we may drop higher-order terms of $q_j$ and only keep the lowest order, $q=q_1+\tilde q$, which constitutes the sub-leading order approximation used in the main text.

%\bibliography{nonherm_reference.bib}
%

\end{document}